# Coulomb Blockade Spectroscopy of a MoS₂ Nanotube

*Simon Reinhardt, Luka Pirker, Christian Bäuml, Maja Remškar,
and Andreas K. Hüttel\**

Low-temperature transport spectroscopy measurements on a quantum dot lithographically defined in a multiwall MoS₂ nanotube are demonstrated. At $T = 300$ mK, clear Coulomb blockade is observed, with charging energies in the range of 1 meV. In single-electron tunneling, discrete conductance resonances are visible at finite bias. Additionally, a magnetic field perpendicular to the nanotube axis reveals clear indications of quantum state transitions, with effective $g$ factors consistent with published theoretical predictions.

When constraining 2D electronic systems further to obtain quantum dots, zero-dimensional electronic systems with discrete quantum states, both generating well-defined boundary conditions and reaching a discrete density of states are significant challenges. Recently, semiconducting 2D transition metal dichalcogenides (TMDCs) have attracted interest as potential material systems for quantum dots. Based on the strong spin-orbit coupling, leading to a spin-split band structure, multiple types of electronic qubits, utilizing spin and valley degrees of freedom, have been proposed.[1–4] Unlike III–V semiconductors, TMDCs in principle allow zero nuclear spin materials, avoiding hyperfine interactions with the electron spin.

From the experimental side, quantum dots have been defined not only in layers of MoS₂,[5–10] but also WSe₂[11] and WS₂.[12] The large electron effective mass, recently found to be $m^* \approx 0.7\,m_e$ for MoS₂[13] and significantly exceeding previous density functional theory (DFT) results,[2,14] leads to a small level spacing. Consequently, so far, it has been a challenge to reach the regime of single quantum level transport in nanofabricated TMDC quantum dots. Discrete-level spectroscopy would allow detailed analysis of the coupled spin/valley physics in the TMDC conduction band.

Here, we present first transport spectroscopy data on a quantum dot defined in a semiconducting multiwall MoS₂ nanotube. Similar to carbon nanotubes[15,16] and semiconducting nanowires,[17,18] the geometry creates a quasi-1D confinement, with intrinsically larger single-particle energy scales. Low-temperature measurements performed at 300 mK are dominated by Coulomb blockade, with regular Coulomb oscillations and features of quantum confinement. In a perpendicular magnetic field, we observe clear indications of quantum state transitions, with effective g factors consistent with published theoretical predictions.

MoS₂ nanotubes are grown by a chemical transport reaction in a two-zone furnace, using iodine as a transport agent.[19,20] The slow, near-equilibrium growth from the vapor phase leads to clean multiwall nanotubes with lengths up to hundreds of micrometers and relatively small density of defects in comparison with other types of growth. As shown in **Figure 1**a,b, the tubes grow individually without forming ropes or bundles. Typical diameters range from ≈10 nm up to several micrometers. The nanotubes are hollow with a wall thickness from 25% to 30% of their diameter. The wall of the MoS₂ nanotube shown in Figure 1c is composed of 25–26 molecular layers. Its diameter is $52 \pm 0.2$ nm. The electron diffraction pattern of Figure 1c reveals that the nanotube has grown in a chiral mode. All walls share a similar angle of chirality, $\alpha \approx 16°$.[21,22] From the growth ampule, the nanotubes are transferred to a highly p-doped silicon wafer with a 500 nm-thick thermally grown surface oxide. The transfer is performed using a wafer dicing tape with low adhesion. This way, a large number of individual nanotubes can be spread over the chip surface and can subsequently be detected using optical microscopy. Scanning electron microscopy (SEM) imaging of nanotubes or the finished devices is not performed to avoid charging of traps in the SiO₂ surface and contamination with hydrocarbons.

Figure 1d shows an optical micrograph of the nanotube used for device 1. Based on previous comparisons of optical and SEM micrographs, we estimate a diameter of 20–50 nm. The device design is shown in Figure 1e. Contacts with a separation of ≈450 nm are defined using standard electron beam lithography, followed by deposition of 30 nm of scandium and 80 nm of gold. To avoid the formation of high Schottky tunnel barriers at the contacts, we use scandium as contact metal because of its low work function.[23,24] Tensile strain in the tube potentially leads

S. Reinhardt, C. Bäuml, Dr. A. K. Hüttel
Institute for Experimental and Applied Physics
University of Regensburg
93040 Regensburg, Germany
E-mail: andreas.huettel@ur.de

L. Pirker, Prof. M. Remškar
Department of Solid State Physics
Institute Jožef Stefan
1000 Ljubljana, Slovenia

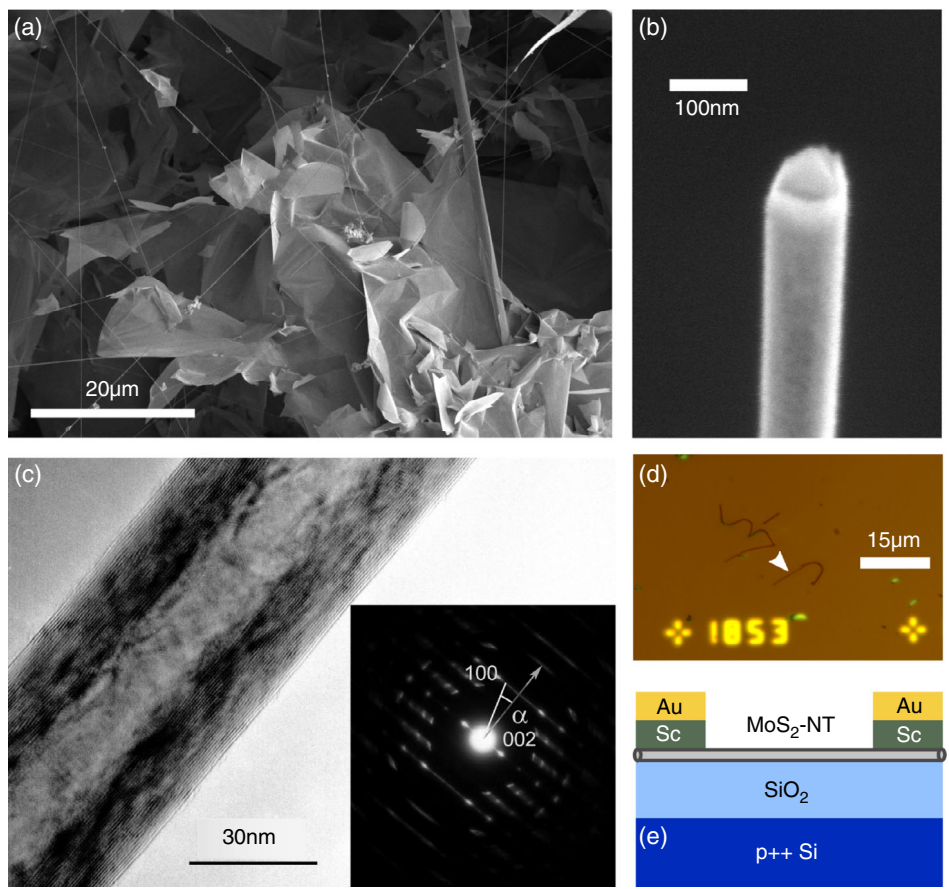

**Figure 1.** a,b) Scanning electron micrographs of $MoS_2$ nanotubes on the growth substrate. Individual nanotubes grow on top of plate-like crystals. c) TEM image of a single $MoS_2$ nanotube with the corresponding diffraction pattern. The arrow marks the nanotube axis. d) Optical microscope image of $MoS_2$ nanotubes transferred to a chip substrate. The arrowhead highlights the nanotube used in the subsequent measurements. e) Device schematic.

to an increase in the $MoS_2$ affinity energy and thus contributes to the further reduction of the Schottky barrier.[25]

At room temperature, the fabricated devices show linear $I(V_{SD})$ characteristics with typical resistances in the order of $\approx 1$ MΩ. When cooled to 300 mK, all studied devices display nonlinear $I(V_{SD})$ behavior, with a suppression of current around zero bias voltage. **Figure 2**a shows a resulting stability diagram. We observe a sequence of Coulomb oscillations with high regularity, indicating the formation of a quantum dot on the contacted tube segment.

In addition, in Figure 2a, a clear threshold bias voltage of $V_{SD} \approx 4$ mV is visible. This indicates that the electronic system consists of a chain of several quantum dots.[26] Unlike the typical shard-like features in a disordered system of quantum dots, here, the onset of neighboring single-electron tunneling (SET) regions is at similar bias voltage; also, only a single set of slopes of the SET region edges is observed, see Figure 2a. This suggests that only one of the quantum dots in the system has a strong capacitive coupling to the back gate; this quantum dot extends over most of the channel length. The additional dots are located near the contact electrodes, where the potential of the back gate is mostly screened. The deposition of reactive metals, as, e.g., scandium, leads to interfacial reactions with $MoS_2$, a likely cause of such additional trap states.[27,28]

The overall conductance data quality is reminiscent of early works on carbon nanotubes, where the macromolecule was (as it is here) in direct contact with a $SiO_2$ substrate surface, its dangling bonds, and disorder. Either suspending the nanotube or encapsulating it into hexagonal boron nitride is thus likely to lead to significantly cleaner and more stable transport spectra.

Analyzing the sequence of Coulomb diamonds highlighted in Figure 2a, we extract an average gate conversion factor, $\alpha_G = 0.048$, and an average gate voltage spacing $\Delta V_G = 23$ mV. From this, we calculate the charging energy, $E_C = e\alpha_G \Delta V_G = 1.1$ meV, and the gate capacitance, $C_G = e/\Delta V_G = 7$ aF. The expected value of the geometric gate capacitance is $C_{G,th} = 2\pi L \epsilon_0 \epsilon_r / \ln(2h/r)$ with the channel length $L$, the tube radius $r$, the thickness of the dielectric layer $h$, and an empirical effective dielectric constant $\epsilon_r = 2.2$.[29] Choosing $L = 450$ nm, $r = 10$ nm, and $h = 500$ nm yields a value of $C_{G,th} = 12$ aF, in good agreement with the measurement, indicating that the size of the quantum dot is defined by the metal electrodes.[30–32]

Figure 2c enlarges the detail marked in Figure 2b with a dashed rectangle. Within the SET region, pronounced discrete conductance resonances corresponding to an excitation energy $\Delta E \approx 500$ μeV are visible. Such a large single-particle excitation energy is a strong indication that we have reached the limit of 1D confinement, where the level spacing is proportional to the results on circular quantum dots in 2D MoS$_2$, predicting both spin and valley effective $g$ factors in this order of magnitude.[2] For the case of MoS$_2$ nanotubes, or generally TMDC nanotubes, further theoretical modeling is needed to uncover the precise electronic structure in strong magnetic fields.

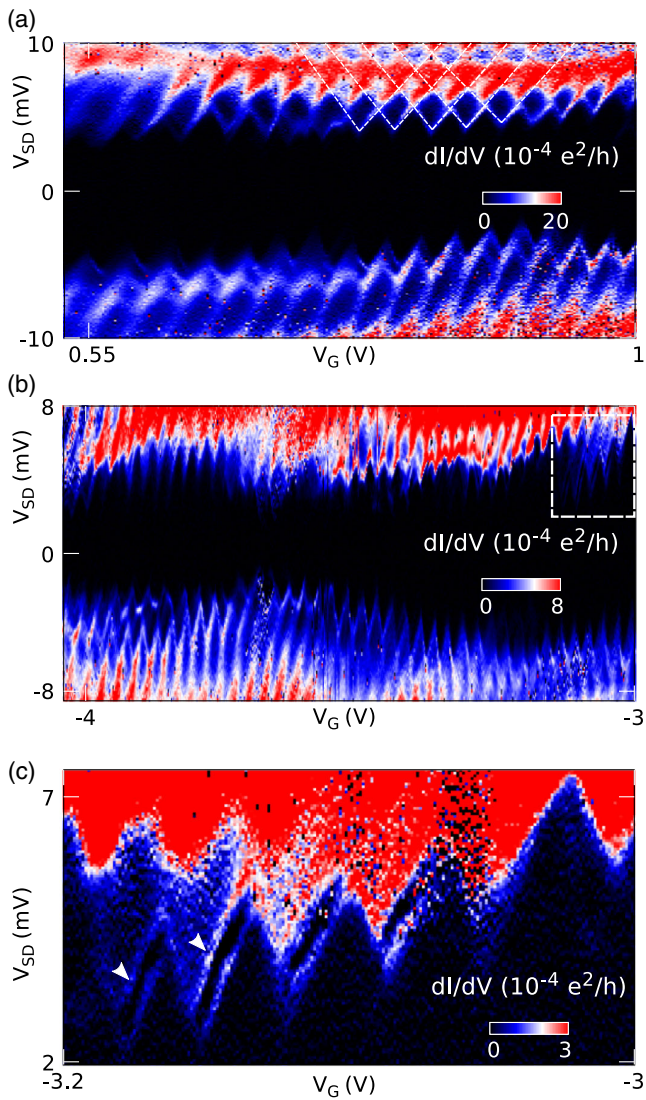

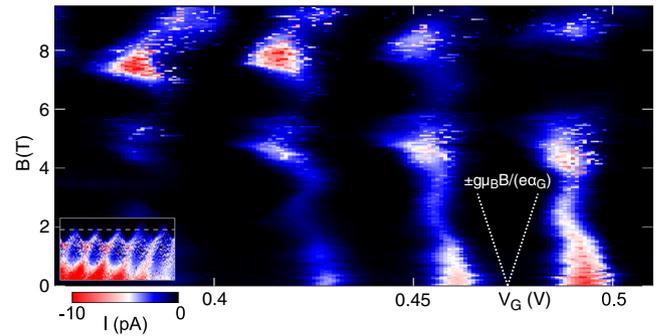

**Figure 3.** $I(V_G, B)$ at constant bias $V_{SD} = -3.5$ mV for $T = 300$ mK. $V_G$ ranges over a sequence of multiple Coulomb oscillations, as indicated in the inset. The dotted lines correspond to the characteristic Zeeman slopes $\partial V_G / \partial B = \pm (g\mu_B)/(e\alpha_G)$ for $g = 2$.

In conclusion, we have performed the first low-temperature transport experiments on a TMDC multiwall nanotube quantum dot. A chemical vapor transport reaction with slow, near-equilibrium growth conditions leads to clean MoS$_2$ nanotubes with very low intrinsic disorder. By using scandium contacts, we avoid the formation of Schottky barriers at the contacts. Current measurements at 300 mK are dominated by Coulomb blockade, with regular Coulomb oscillations and clear resonant features in nonlinear transport overlaid on a low-bias region of suppressed conductance. In a perpendicular magnetic field, we observe clear indications of quantum state transitions, with effective $g$ factors consistent with published calculations.[2]

The low-bias suppression of conductance is a typical feature of an array of multiple quantum dots; however, the measurement shows that this disorder is limited to the contact regions. Future work shall improve the contact properties, toward detailed single quantum dot level spectroscopy in aligned magnetic fields and exploration of the confinement spectrum in the band structure of the material. An obvious next step is also to encapsulate or suspend the nanotubes, in order to suppress substrate charge influences. Furthermore, WS$_2$ nanotubes have already been shown to exhibit intrinsic superconductivity at strong doping,[33,34] pointing toward the possibility of intrinsic 1D semiconductor–superconductor heterostructures within a single macromolecule.

**Figure 2.** Stability diagrams measured at $T = 300$ mK: differential conductance d$I$/d$V_{SD}$ ($V_{SD}$, $V_G$) obtained by numerical differentiation of the measured dc current $I(V_{SD}, V_G)$. For better contrast, the conductance scale has been cut off as indicated. a) Overview measurement of the region $0.5$ V $\leq V_G \leq 1$ V, where the highlighted edges of SET onset have been used to estimate charging energy and gate capacitance. b) Overview measurement for $-4$ V $\leq V_G \leq -3$ V, illustrating the slow modulation of the conduction threshold with gate voltage. c) Detail plot of the highlighted region in (b), displaying discrete conductance resonances within the SET regions.

number of charges on the quantum dot, similar to measurements on semiconducting nanowires.[30]

Finally, we have collected first data on the influence of a perpendicular magnetic field on the current through the quantum dot. **Figure 3** shows the measurement of $I(V_G, B)$ over a sequence of four Coulomb oscillations at a fixed finite bias voltage, $V_{SD} = -3.5$ mV, tracing across the "tips" of the SET regions, as indicated in the inset of Figure 3. The conductance maxima shift gate voltage position, with slopes comparable to Zeeman phenomena at $g \approx 2$ and multiple apparent quantum state transitions. This is in good agreement with DFT-based

## Acknowledgements

The authors acknowledge financial support by the Deutsche Forschungsgemeinschaft via SFB 1277, the DAAD via the PPP Slovenia program, grant no. 57401796, and the Slovenian Research Agency. The authors thank Ch. Strunk and D. Weiss for the use of experimental facilities. The data have been recorded using the Lab:Measurement software package.[35]

## Conflict of Interest

The authors declare no conflict of interest.

## Keywords